\documentclass{rmaa}

\usepackage{paralist}
\usepackage{psfrag,color}

\title{A model for the cross section of a turbulent, radiative jet or wake} 

\author{
  J. Cant\'o,\altaffilmark{1}
  A. C. Raga,\altaffilmark{2} 
  and A. Riera\altaffilmark{3,4} \medskip}

\altaffiltext{1}{Instituto de Astronom\'\i a, Univer\-si\-dad
Nacio\-nal Aut\'o\-noma de M\'e\-xico, M\'exico.}
\altaffiltext{2}{Instituto de Ciencias Nucleares,
 Univer\-si\-dad
Nacio\-nal Aut\'o\-noma de M\'e\-xico, M\'exico.}
\altaffiltext{3}{Departament de F\'\i sica i Enginyeria
Nuclear, Universitat Polit\`ecnica de Catalunya,
Spain.}
\altaffiltext{4}{On sabbatical leave at the ICN-UNAM.} 

\shortauthor{Cant\'o, Raga, \& Riera}
\shorttitle{Turbulent jet/wake}

\fulladdresses{
\item Jorge Cant\'o: Instituto de Astronom\'\i a, UNAM, Apdo. Postal 70-264,
04510 M\'exico, D. F., M\'exico.
\item Alejandro C. Raga: Instituto de Ciencias Nucleares, UNAM,
 Apdo. Postal 70-543, 04510 M\'exico, D. F., M\'exico. 
  (raga@nuclecu.unam.mx).
\item Angels Riera: Departament de F\'\i sica i Enginyeria
Nuclear, Universitat Polit\`ecnica de Catalunya, Av. V\'\i ctor
Balaguer s/n, E-08800 Vilanova i la Geltr\'u, Spain.
  (angels.riera@upc.es). }

\listofauthors{J. Cant\'o, A. C. Raga, \& A. Riera}
\indexauthor{Cant\'o, J.}
\indexauthor{Raga, A. C.}
\indexauthor{Riera, A.}
\SetVolume{39} \SetFirstPage{000} \SetYear{2003}

\ReceivedDate{2003 February 10} \AcceptedDate{2003 May 14}

\abstract{We present an analytical model for the cross section
of a turbulent, radiative jet or wake. This model is appropriate
for modeling HH jets, or ``wakes'' left behind by ``astrophysical
bullets''. Even though the model is very simple, it has the
benign property of only having four free parameters (the outer
radius of the beam, the axial
velocity, the velocity at the edge of the beam, and the turbulent
velocity width), which can be derived by fitting the radial velocity
and line width cross sections of an observed outflow. We illustrate
how to do such fits using previously published spectroscopic
data of the HH~110 jet.}

\resumen{Presentamos un modelo anal\'\i tico de la secci\'on
de un flujo turbulento y radiativo. Este modelo es apropiado
para modelar chorros HH o ``estelas'' detr\'as de ``balas
astrof\'\i sicas''. A pesar de que el modelo es muy simple,
tiene la propiedad benigna de tener s\'olo cuatro par\'ametros
libres (el radio exterior del flujo,
la velocidad axial, la velocidad del borde del haz, y la
dispersi\'on de velocidades turbulentas), que pueden ser derivados
mediante una comparaci\'on con las secciones de la velocidad radial
y del ancho de l\'\i nea de un objeto observado. Ilustramos como
realizar este tipo de ajustes usando observaciones espectrosc\'opicas
previamente publicadas del chorro HH~110.}

\addkeyword{ISM: Herbig-Haro objects}
\addkeyword{ISM: individual (HH~110)}
\addkeyword{ISM: Jets and outflows}
\addkeyword{ISM: kinematics and dynamics}

\RescaleTitleLengths{0.9}
\begin{document}

\maketitle

\section{Introduction}

Some HH jets show complex structures of emitting knots that are
reminiscent of turbulent, laboratory jets. Two examples of this
kind of flow are HH~111 (Reipurth, Raga, \& Heathcote 1996) and HH~399
(Cernicharo  et~al.\@  1998; Rosado  et~al.\@  1999). The morphology of
these jets resembles the radio continuum maps of Faranoff-Riley
Type~I extragalactic jets, which have been modeled in terms of
analytical ``mean flow+turbulent eddy'' fully turbulent jet models
(e.g., Bicknell 1984, 1986; Komissarov 1988, 1994).

For HH jets, some effort has been done to model the turbulent
boundary layer around a laminar jet beam core (e.g., Cant\'o
\& Raga 1991; Lim, Rawlings, \& Williams 1999; Binette et~al.\@ 1999).
However, models for a fully turbulent jet have been quite
primitive (Richer, Hills, \& Padman 1992; Raga  et~al.\@  1993), and limited
to a description of the general dynamical properties of such
a jet as it incorporates mass from the surrounding environment.
Also, 2 and 3D numerical simulations of the development of turbulence
in radiative jets have been carried out (Massaglia  et~al.\@  1996;
Rossi  et~al.\@  1997; Downes \& Ray 1998;
Stone, Xu, \& Hardee 1997; Xu, Hardee, \& Stone 2000; Micono  et~al.\@  2000).

In the present paper, we discuss an analytical model
for the cross section of a radiative, turbulent jet. Even though
the model is dynamically very simple, it is useful in that
it leads to concrete predictions of the observational properties
(radial velocity, line widths and line profiles) that should
characterize the cross section of a turbulent jet. The usefulness
of this model is then illustrated by carrying out a comparison
with long-slit spectra of the HH~110 jet (taken from Riera  et~al.\@  2003a).

We present the dynamical model in \S~2. The derivation of
the line profiles and their moments (barycenter and line
width) is discussed in \S~3. The comparison with observations
of HH~110 is made in \S~4. Finally, the results are discussed
in \S~5.

\section{A simple model for a turbulent jet or wake}

It is common practice to describe a turbulent flow as a superposition
of a ``mean flow'' (corresponding in principle to an ensemble average
of many ``experiments'', but which can also be estimated by appropriately
defined spatial or temporal averages) and highly time-dependent
and chaotic ``turbulent eddies''. It is well known that both turbulent
laboratory jets and turbulent wakes have an axially peaked mean flow
velocity, which is directed mainly along the flow axis.
Even though there are no experiments of radiative, high Mach number jets,
3D numerical simulations of such jets (e.g., Micono  et~al.\@  2000)
appear to show that when such flows become turbulent, they also develop
an axially peaked mean flow velocity profile.

In order to develop an
analytical model of the cross section of the jet, we will then consider
a Taylor series expansion $v_j(r)=a+br+cr^2+...$ for the mean flow
velocity (which we assume to be directed along the flow axis).
It can be argued that in order not to have an unphysical axial ``cusp'',
the first order term has to be equal to zero. Therefore, the lowest order,
physically meaningful, expansion is quadratic. We then consider the
simplest possible form for the mean flow velocity cross section:

\begin{equation}
v_j(r)=v_0\,\left(1-{r^2\over h^2}\right)+v_1\,,
\label{vj}
\end{equation}
where $h$ is the outer radius of the jet beam, $v_0+v_1$ is the
axial velocity, and $v_1$ is the velocity of the material in the outer
edge of the jet beam (with $r=h$). We will assume that $h$, $v_0$, and
$v_1$ vary only slowly along the beam of the jet (in other words, that
they change over scales much larger than $h$), so that they can be
considered as constants when studying the properties of the jet cross
section.

Superimposed on this mean velocity, one also has the turbulent eddies.
We will assume that these motions are randomly directed, and that
the component of this velocity along any direction has a Gaussian
probability distribution with a mean value of zero, and a dispersion
$\Delta v_T$ which is independent of position. For subsonic flows,
it is normal to set $\Delta v_T \propto v_j$. However, for supersonic
flows it appears that laboratory experiments are consistent
with a $\Delta v_T \propto c_s$ assumption (where $c_s$ is the sound speed,
---see Cant\'o \& Raga 1991). As a radiative turbulent jet is approximately
isothermal (see below), this leads to the conclusion that $\Delta v_T$
should be independent of position.

As is normally done in models of turbulent jets, we will assume that
the jet beam is in lateral pressure equilibrium with the surrounding
material. Also, it has been previously argued (Cant\'o \& Raga 1991;
Raga  et~al.\@  1993) that a radiative, turbulent flow reaches a local
balance between the turbulent dissipation and the radiative energy loss,
and that because of the steepness of the cooling function this balance
always leads to a temperature of a few thousand K. Therefore, the flow
is approximately isothermal. Together with the pressure balance condition,
this leads to the conclusion that the density of the jet has to be approximately
constant across the beam of the jet.

If the temperature and density across the section of the jet are constant,
it then appears to be reasonable to assume that
the emission coefficient $j$ associated with a given emission line is
independent of position across the section of the jet. However, this
is not necessarily true for the emission lines which are responsible
for the radiative cooling (e.g., the [O~I], [O~II], [C~II]
collisionally excited lines), which actually force the near isothermality
through their strong temperature dependence. The assumption of a
position-independent $j$ is therefore only reasonable for lines such
as the recombination lines of H, which have a temperature dependence
which is much shallower than the one of the forbidden lines which
dominate the radiative cooling of the gas. In the following, we also
assume that the emission line under consideration (which could be,
 e.g., H$\alpha$) is optically thin.

In this way, one can construct what is basically the simplest possible
model for the mean flow cross section of a radiative turbulent
jet or of a wake (which are completely equivalent at the level of
approximation of the present model). This flow has a quadratic
velocity profile (see eq.~\ref{vj}), and a position-independent
line emission coefficient and turbulent velocity dispersion $\Delta v_T$.
From this simple structure, one can then compute the emitted line profiles.
This is done in the following section.

\begin{figure}[!t]
  \includegraphics[width=\columnwidth]{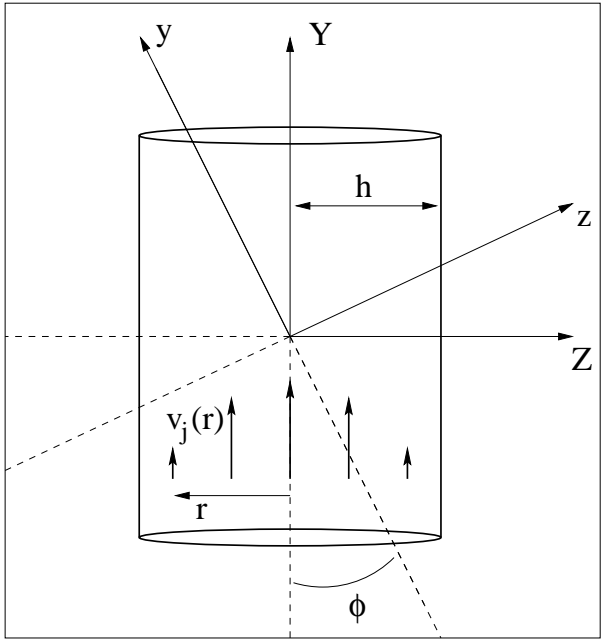}
  \caption{Schematic diagram showing a turbulent jet with a mean flow
velocity $v_j$ that decreases monotonically as a function of the
cylindrical radius $r$. The jet travels along the $Y$-axis.
The $X$-axis, which is parallel to the plane of the sky, points out
of the page, directly towards the reader. A second coordinate system
is defined, with the $z$-axis along the line of sight, and the $y$-axis
on the projection on the plane of the sky of the jet axis. The jet
is shown moving at a (negative) angle $\phi$ (with respect to the
plane of the sky) towards the observer.}
  \label{fig1}
\end{figure}

\section{The emission line profiles}

Let us consider an observation of a jet which is moving at an angle
$\phi$ with respect to the plane of the sky (with negative values of
$\phi$ corresponding to motions towards the observer). One can construct
an orthogonal coordinate system $(X,Y,Z)$, with the jet moving along
the $Y$-axis and with the $X$-axis on the plane of the sky. The observer
defines a second, $(x,y,z)$ coordinate system, with $y$ along the
projection on the plane of the sky of the jet axis, $x$ across the
projected cross section of the jet, and $z$ along the line of
sight. These two coordinate systems are shown in the schematic
diagram of Figure~1.

The component along $z$ of the mean flow velocity (i.e., the radial
velocity associated with the mean flow) then is:

\begin{equation}
v_m(x,z)=v_0\sin\phi\,\left(1-{{x^2+z^2\cos^2\phi}\over h^2}\right)
+ v_1\sin\phi\,,
\label{vm}
\end{equation}
as can be deduced from eq.~(\ref{vj}) and simple geometric relations
between the $(X,Y,Z)$ and $(x,y,z)$ coordinates.

The radial velocity-dependent emission line coefficient is then given by
\begin{equation}
j_v(x,z)=j_0\,\Psi_v(x,z)\,,
\label{jv}
\end{equation}
where $j_0$ is constant (see \S~2) and
\begin{equation}
\Psi_v={1\over \sqrt{\pi}\Delta v}e^{-(v-v_m)^2/\Delta v^2}\,,
\label{psi}
\end{equation}
where $v_m$ is given by eq.~(\ref{vm}) and the velocity
disperion $\Delta v$ includes both the turbulent and the thermal
motions of the gas. Also, in order to compare predictions of the model
with observations, one can include the instrumental broadening
by adding it in quadrature to the thermal+turbulent line width.

The specific intensity along a line of sight passing at a distance
$x$ from the projected outflow axis is then given by
\begin{equation}
I_v(x)=2\int_0^{z_m}j_0\Psi_v dz\,,
\label{iv}
\end{equation}
where
\begin{equation}
z_m={{\left(h^2-x^2\right)^{1/2}}\over \cos \phi}\,.
\label{zm}
\end{equation}
For non-zero $\Delta v$, the integral of eq.~(\ref{iv}) cannot
be performed analytically. In order to calculate it numerically,
we can write it in dimensionless form:
\begin{equation}
I_\nu={1\over \Delta \nu}
\int_0^1 e^{-[\nu-(1-\eta^2)]^2/\Delta \nu^2}\,d\eta\,,
\label{inu}
\end{equation}
where $\eta=z/z_m$ and
\begin{equation}
I_\nu={\sqrt{\pi}v_0\sin\phi\cos^2\phi\, {z_m}^2\over 2j_0h^2}I_v\,,
\end{equation}
\begin{equation}
\nu={{h^2\left(v-v_1\sin\phi\right)}\over v_0\sin\phi\cos^2\phi\, {z_m}^2}\,,
\label{nu}
\end{equation}
\begin{equation}
\Delta \nu={h^2\Delta v\over v_0\sin\phi\cos^2\phi\, {z_m}^2}\,.
\label{dnu}
\end{equation}
In the $\Delta \nu\to 0$ limit, the integral of eq.~(\ref{inu})
can be performed analytically, giving
\begin{equation}
I_\nu\to {\sqrt{\pi}\over 2\sqrt{1-\nu}}\,.
\label{ivan}
\end{equation}
For nonzero $\Delta \nu$ one can
compute the dimensionless line profiles by carrying
out a numerical integration of eq.~(\ref{inu}). Alternatively,
one could compute the same line profile by convolving
eq.~(\ref{ivan}) with a Gaussian of dispersion $\Delta \nu$ (which
one can prove is mathematically equivalent to eq.~\ref{inu}).
The results of such integrations are shown in Figure~2.

\begin{figure}[!t]
  \includegraphics[width=\columnwidth]{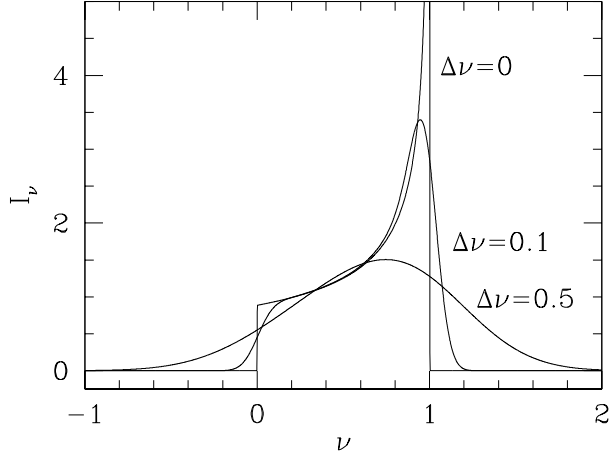}
  \caption{Dimensionless line profiles $I_\nu$
 (see eq.~\ref{inu}) as a function of the dimensionless radial velocity
$\nu$ (see eq.~\ref{nu}) for three different values of the
dimensionless turbulent+thermal line width $\Delta \nu$ (see
eq.~\ref{dnu}). The solution for $\Delta \nu=0$ is given
analytically by eq.~(\ref{ivan}).}
  \label{fig2}
\end{figure}

We now compute three different velocity-moments of the line
profile (given by eq.~\ref{iv}). The line flux is:
\begin{equation}
I_0\equiv\int_{-\infty}^\infty I_v(x)\,dv=2j_0z_m=2j_0
{{\left(h^2-x^2\right)^{1/2}}\over \cos\phi}\,.
\label{i0}
\end{equation}
The barycenter of the line profile is given by:
$$V_c\equiv {1\over I_0}\int_{-\infty}^\infty v\,I_v(x)\,dv$$
\begin{equation}
={2\over 3}v_0\sin\phi{{\left(h^2-x^2\right)}\over h^2}+v_1\sin\phi\,.
\label{vc}
\end{equation}
Finally, the velocity dispersion of the line profile is:
$$W^2\equiv {1\over I_0}\int_{-\infty}^\infty (v-V_c)^2\,I_v(x)\,dv$$
\begin{equation}
={\Delta v^2\over 2}+{4\over 45}{v_0}^2\sin^2\phi
\left(1-{x^2\over h^2}\right)^2\,.
\label{w}
\end{equation}

These expressions for the line center (eq.~\ref{vc}) and for
the line width (eq.~\ref{w}), as well as the shape of the line
profile itself (eq.~\ref{iv}), can in principle be compared directly
with observations of turbulent, radiative astrophysical jets. In the
following section, we show such a comparison between our model and
previously published observations of HH~110.

\begin{figure}[!t]
  \includegraphics[width=\columnwidth]{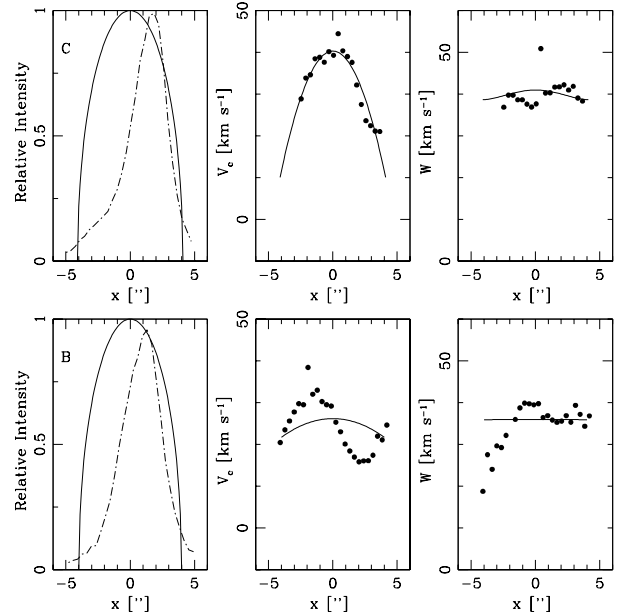}
  \caption{H$\alpha$ intensity (dashed line, left), barycenters (dots,
centre) and line widths (dots, right) as a function
of position across the width of the HH~110 jet, obtained from the
long-slit spectra of Riera  et~al.\@  (2003a). The results obtained
for slit positions across knot B (bottom) and knot C (top) are shown.
The abscissa gives the position as offsets from the outflow axis.
The observed radial velocities are given relative to the
velocity of the nearby molecular cloud, and are plotted as
positive values (the HH~110 flow, however, is blue-shifted).
The solid lines show the results from the model fits described
in \S~3 (see also Table~1).}
  \label{fig3}
\end{figure}

\section{A comparison with observations \linebreak of HH~110}

Riera  et~al.\@  (2003a) obtained high resolution, long-slit spectra
across knots B and C of the HH~110 jet (which lie at
distances of $\approx 8$ and $23''$ down the jet axis from knot
A). From the resulting
H$\alpha$ position-velocity (PV) diagrams, we can compute the barycenter
and width of the line profiles (using the first equalities of
eqs.~\ref{vc} and \ref{w}, respectively) as a function of
position across the jet axis.

The obtained results are shown in Figure~3.
From this figure, we see that for knot C one obtains a centrally
peaked $V_c(x)$ cross section, and a flat $W(x)$ dependence.
For knot B, one obtains a more complex, asymmetric $V_c(x)$
cross section.

We now proceed as follows. We fix the width of the jet to a value
$h=4''$ (which approximately corresponds to the half-width of the
region with observed emission). Then, we carry out a least
squares fit of the $V_c(x)$ dependence predicted from the model
(eq.~\ref{vc}) in order to obtain the values of $v_0\sin\phi$
and $v_1\sin\phi$ which best fit the data. Finally, using these
values for $h$, $v_0\sin\phi$ and $v_1\sin\phi$, we carry out
a fit to the observed line widths with the $W(x)$ predicted from
the model (eq.~\ref{w}) in order to determine the
value of $\Delta v$. In this way, we determine the parameters of
the model that best fit the observed cross sections of the HH~110
jet.

\begin{table}[!t]\centering
  \newcommand{\DS}{\hspace{6\tabcolsep}} 
  \setlength{\tabnotewidth}{\linewidth}
  \setlength{\tabcolsep}{1.33\tabcolsep}
  \tablecols{6}
  \caption{Model fits to HH 110\tabnotemark{a}}
  \begin{tabular}{cccccc}
    \toprule
Knot & h & $v_0\sin\phi$ & $v_1\sin\phi$ & $v_0+v_1$\tabnotemark{b} &
$\Delta v$ \\
 & [$''$] & \multicolumn{4}{c}{\hfil [km~s$^{-1}$] \hfil}\\
\midrule

B & 4 & \phantom{0}7 & 22 & 50 & 51 \\
C & 4 & 45 & 10 & 97 & 55 \\
    \bottomrule
    \tabnotetext{a}{Fits to the long-slit spectra across knots
B and C ob\-tained by Riera  et~al.\@  (2003a).}
    \tabnotetext{b}{Computed for an orientation angle $\phi=-35^\circ$
with re\-spect to the plane of the sky.}
  \end{tabular}
\end{table}

The model parameters resulting from these fits are given
in Table~1. For knot B, the least squares fit gives a flat
$V_c(x)$ cross section, which does not reproduce well the
rather complex, observed cross section (see Figure~3). On the
other hand, a more convincing fit is obtained for the
$V_c(x)$ cross section of knot C. For both knots B and C, the
observed H$\alpha$ intensity cross sections are much sharper and more
asymmetric than the broader, symmetric intensity cross section predicted
from the model.

The model fits show an outward ``acceleration'', with a higher
axial velocity for knot C than for knot B. This result is in
agreement with the acceleration down the jet axis noted by
Riera  et~al.\@  (2003a). On the other hand, the model fits to both
knots give similar, $\Delta v\sim 50$~km~s$^{-1}$ line broadenings
(see eq.~\ref{psi} and Table~1). This line broadening clearly
exceeds the 20~km~s$^{-1}$ instrumental broadening of the data of
Riera  et~al.\@  (2003a), and therefore mostly reflects the turbulent
motions of the gas flowing down the HH~110 flow.

\begin{figure}[!t]
  \includegraphics[width=\columnwidth]{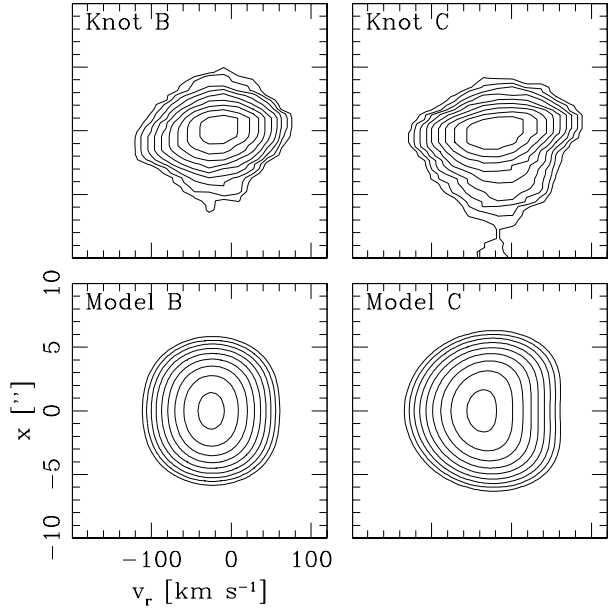}
  \caption{Observed (top) and predicted (bottom) H$\alpha$
PV diagrams for knots B (left) and C (right). The radial velocities
of the observed PV diagrams are given with respect to the radial
velocity of the nearby molecular cloud. The positions are
given as offsets across the jet beam, measured
from the outflow axis. The PV diagrams are depicted
with logarithmic, $\sqrt{2}$ contours.}
  \label{fig4}
\end{figure}

Using these model fits, we can now calculate predicted PV
diagrams (for a slit position across the jet axis) by numerically
integrating eq.~(\ref{iv}). We  then convolve the
model predictions with a Gaussian ``seeing'' of FWHM$=2''.5$ in
order to simulate the conditions of the observations of
Riera  et~al.\@  (2003a). The results of this exercise are shown in
Figure~4, together with the observed H$\alpha$ PV
diagrams for knots B and C.

Comparing the predicted and observed PV diagrams,
one sees that while their general characteristics (e.g., the
radial velocities, characteristic line widths and spatial extension 
of the emission) agree well (as guaranteed by the model fits described
above), they do have substantial differences. In particular, the
observed PV diagrams show clear asymmetries on the two sides of
the outflow axis, which are of course absent in the axisymmetric
theoretical model.

\section{Conclusions}

We have developed a simple, analytical model that describes the
cross section of a radiative, turbulent flow. This flow could
correspond to a turbulent jet beam, or to a turbulent wake left
behind by the passage of an ``astrophysical bullet''.

Though the model is extremely simple, it has the redeeming property
that it leads to clear, analytic predictions of the spatial
dependence (across the outflow) of the emission line profiles.
In particular, we derive relations giving the barycenter and
the line width as a function of position across the outflow.

We have then shown how these relations can be used to model
the observed cross sections of the HH~110 jet (taken at two
different positions along the jet beam). From fits to the
observed barycenter and line widths, one can derive the four
free parameters of the model~: the outer radius of the beam,
the central velocity, the velocity
of the outer edge of the beam and the ``turbulent width'' of the
emitted lines.

Interestingly, we find that while the observed radial
velocity cross section of knot~C (of HH~110) does resemble the
functional form predicted by our model, the cross section of
knot~B does not (see Fig.~3). This kind of result is not
surprising, since a truly turbulent flow will have eddies of
all sizes (up to sizes comparable to the width of the jet).
Therefore, comparisons of observed cross sections (taken at a given
position along the jet) with predictions from a ``mean flow model''
(in which the turbulent eddies are only considered in the form
of a line broadening) are likely to be unsatisfactory.

In order to obtain more appropriate comparisons between our analytical
model and observations of HH jets it would be necessary to have
spectroscopic observations with 2D spatial resolution. With such
observations, one could calculate jet cross sections from the
emission averaged over an appropriately defined length along the outflow
axis. This ``averaging length'' should be larger than a few jet diameters,
so that an average over several characteristic sizes of
the turbulent eddies is carried out.

In this way, one could try
to recover the properties of the ``mean flow'' from observations
of a turbulent jet or wake. With such results, a more appropriate
comparison with the model described above could be carried out.

This has now been attempted by Riera  et~al.\@  (2003b), who have obtained
Fabry-P\'erot observations of HH~110, and computed cross sections
of the outflow carrying out averages along the jet over the sizes of
the different emission structures (using a wavelet analysis technique).
These authors find that such ``average cross sections'' compare surprisingly
well with the cross sections predicted by the analytic model described
above. This somewhat surprising result appears to indicate that our simple
model does capture some of the important properties of the HH~110 flow.

\medskip

\acknowledgements
The work of JC and ACR was supported by CONACyT grants 36572-E and
34566-E. A. Riera acknowledges the ICN-UNAM for support during her \linebreak

\adjustfinalcols
\noindent
sabbatical. The work of A. Riera was supported by the MCyT grant
AYA2002-00205 (Spain). We acknowledge an anonymous referee for
helpful comments.

\vspace*{\baselineskip}

\end{document}